\pdfoutput=1

\documentclass[11pt]{article}

\usepackage{acl}

\usepackage{times}
\usepackage{latexsym}
\usepackage[T1]{fontenc}
\usepackage[utf8]{inputenc}
\usepackage{microtype}

\usepackage{url}
\usepackage{graphicx}  
\usepackage{amsfonts}
\usepackage{CJK}
\usepackage{bm}
\usepackage{mathrsfs}
\usepackage{float}
\usepackage{xcolor,colortbl}
\usepackage{adjustbox}
\usepackage{subfigure}
\usepackage{booktabs,multirow}
\usepackage{bigstrut}
\usepackage{paralist}
\usepackage{bbm}
\usepackage{makecell}
\usepackage{nicefrac}       
\usepackage{microtype} 
\usepackage{epsfig}
\usepackage{diagbox}
\usepackage{framed}
\usepackage{empheq}
\usepackage{array}
\usepackage{enumitem}
\usepackage{mdwlist}
\usepackage{xspace,mfirstuc,tabulary}
\usepackage{algorithm}
\usepackage{placeins}
\usepackage{algpseudocode}
\usepackage{microtype}

\usepackage[normalem]{ulem}
\useunder{\uline}{\ul}{}
\usepackage{threeparttable}
\usepackage{makecell}
\usepackage{booktabs}

\usepackage{xspace}
\usepackage{natbib}
\usepackage{hyperref}
\usepackage{url}

\usepackage{graphicx}
\usepackage{subfigure}
\usepackage{multirow}
\usepackage{multicol}
\usepackage{pifont}
\usepackage{enumitem}
\usepackage{amsmath}
\usepackage{appendix}

\usepackage[figuresright]{rotating}
\usepackage[most]{tcolorbox}
\newtcbox{\mybox}[1][red]
  {on line, arc = 0pt, outer arc = 0pt,
    colback = #1!10!white, colframe = #1!50!black,
    boxsep = 0pt, left = 1pt, right = 1pt, top = 2pt, bottom = 2pt,
    boxrule = 0pt, bottomrule = 1pt, toprule = 1pt}
\definecolor{BoxBackground}{RGB}{240, 240, 240} 
\definecolor{BoxFrame}{RGB}{0, 0, 0} 
\definecolor{TitleBackground}{RGB}{0, 0, 0} 
\definecolor{TitleText}{RGB}{255, 255, 255} 
\tcbset{
  academicbox/.style={
    boxsep=5pt,
    left=2pt,
    right=2pt,
    bottom=0.5pt,
    boxrule=0.5pt,
    colback=BoxBackground,
    colframe=BoxFrame,
    colbacktitle=TitleBackground,
    coltitle=TitleText,
    enhanced,
    attach boxed title to top left={yshift=-0.1in,xshift=0.1in},
    boxed title style={boxrule=0pt,colframe=white},
    title={#1},
  }
}
\newtcolorbox{AcademicBox}[1][]{academicbox=#1}

\usepackage{tikz}
\usepackage{graphicx}

\def\method{{\sc LongCodeU}\xspace}
\def\textbfmethod{{\sc \textbf{LongCodeU}}\xspace}

\newcommand{\eg}{{\emph{e.g.,}}\xspace}
\newcommand{\ie}{{\emph{i.e.,}}\xspace}

\usepackage{wasysym}

\title{\textbfmethod: Benchmarking Long-Context Language Models on \newline Long Code Understanding}

\author{Jia Li
 \\
  Affiliation / Address line 1 \\
  Affiliation / Address line 2 \\
  Affiliation / Address line 3 \\
  \texttt{email@domain} \\\And
  Xuyuan Guo \\
  Affiliation / Address line 1 \\
  Affiliation / Address line 2 \\
  Affiliation / Address line 3 \\
  \texttt{email@domain} \\}

  \author{Jia Li, \ Xuyuan Guo, \ Lei Li, \ Kechi Zhang, \ Ge Li\footnotemark[2], \textbf{Jia Li\male,} \ \\   \ \textbf{Zhengwei Tao,}  \ \textbf{Fang Liu,} \ \textbf{Chongyang Tao,}  \ \textbf{Yuqi Zhu,} \ \textbf{Zhi Jin\footnotemark[2]} \\
Key Lab of High Confidence Software Technology (Peking University), MoE, \\
School of Computer Science, Peking University, China \\
\texttt{\{lijiaa,zhangkechi\}@pku.edu.cn}, \\ \texttt{lige@pku.edu.cn}, \\ \texttt{zhijin@pku.edu.cn}}

\begin{document}

\maketitle
\renewcommand{\thefootnote}{\fnsymbol{footnote}}
\footnotetext[2]{Corresponding authors.}
\renewcommand{\thefootnote}{\arabic{footnote}}

\begin{abstract}

Current advanced long-context language models offer great potential for real-world software engineering applications. However, progress in this critical domain remains hampered by a fundamental limitation: the absence of a rigorous evaluation framework for long code understanding. To gap this obstacle, we propose a long code understanding benchmark \method from four aspects (8 tasks) to evaluate LCLMs' long code understanding ability required for practical applications, including code unit perception, intra-code unit understanding, inter-code unit relation understanding, and long code documentation understanding. We evaluate 9 popular LCLMs on \method (\ie 6 general models and 3 code models). Our experimental results reveal key limitations in current LCLMs’ capabilities for long code understanding. Particularly, the performance of LCLMs drops dramatically when the long code length is greater than 32K, falling far short of their claimed 128K$\sim$1M context windows. In the four aspects, inter-code unit relation understanding is the most challenging for LCLMs.
Our study provides valuable insights for optimizing LCLMs and driving advancements in software engineering. 
 
\end{abstract}

\section{Introduction}
\label{Introduction}

\begin{figure}[t]
\centering
\setlength{\abovecaptionskip}{0.1cm}
\setlength{\belowcaptionskip}{-5mm}
\includegraphics[width=\columnwidth]{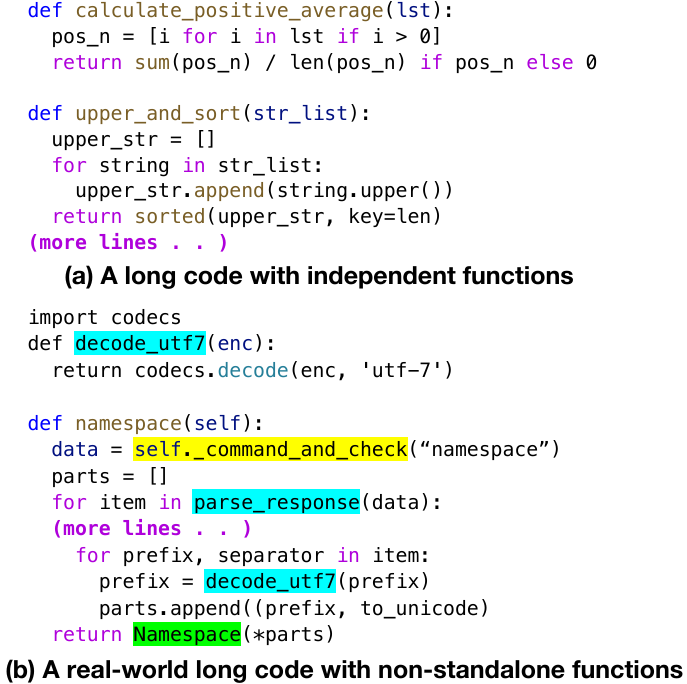}
\caption{Examples of a synthetic long code with independent functions and a real-world long code with non-standalone functions. Dependencies are highlighted.}
\label{figure: benchmark comprison}
\end{figure}

Recent advances in long-context language models (LCLMs) like Gemini-1.5 \cite{team2024gemini} and GPT-4o \cite{GPT-4o}, which support context windows exceeding hundreds of thousands of tokens, offer unprecedented potential for real-world software engineering applications. These models promise transformative improvements in related downstream tasks, such as repository-level code generation \cite{zhang2024codeagent, bi2024iterative}, real-world GitHub issues resolution\cite{jimenez2023swe}, and long code summarization \cite{dhulshette2025hierarchical}. However, progress in this critical area remains hampered by a fundamental limitation: \textbf{the lack of rigorous evaluation frameworks for long code understanding}—a capability essential for real-world software development tools that require accurate code unit perception, intra-code unit understanding, inter-code unit relation understanding, and long code documentation understanding in code repositories.

Current benchmarks generally fall into two categories, which face five fundamental limitations that hinder comprehensive evaluation of LCLMs' long code understanding capabilities.  
The first category includes studies such as RepoQA \cite{liu2024repoqa}, whose task design has insufficient diversity like only focusing on needle function search. While these evaluations are useful, \ding{182} \textbf{they do not capture the real-world full range of code understanding capabilities} needed for practical coding scenarios. Additionally, benchmarks like L-Eval \citet{an2023eval} further compound these limitations by using synthetic ``long code'' through a simple joining of independent code snippets. \ding{183} This approach \textbf{overlooks the natural dependencies between code segments}, as shown in Figure~\ref{figure: benchmark comprison}. 
\ding{184} These studies also face issues with \textbf{data contamination}. They neither enforce temporal constraints on code release dates nor address potential model pre-exposure through evaluating on previously published datasets. \ding{185} Their maximum supported context length of 36.5K tokens also \textbf{falls far short of stress-testing the claimed 128K-1M context windows of modern LCLMs}.
The second category includes benchmarks like Long Code Arena \cite{bogomolov2024long}, LongBench \cite{bai2023longbench}, SWE-bench \cite{jimenez2023swe}, and DevEval \cite{li2024deveval}, which evaluate long-context understanding based on performance in downstream tasks. 
\ding{186} However, these benchmarks \textbf{entangle code understanding with other task-specific challenges} like code generation and bug fixing.
This makes it hard to determine whether performance limitations are due to a lack of code understanding or other factors.

To address these limitations, we propose \method, a benchmark designed to isolate and comprehensively evaluate LCLMs' capacity to understand and reason about real-world, dependency-rich, long code contexts. Our benchmark offers the following key features:
\begin{itemize}[topsep=0pt,noitemsep] 
\item[$\bullet$] \textbf{Comprehensive Tasks stem from Practical Applications.} We evaluate LCLMs from four aspects (8 tasks) to evaluate the LCLMs' long code understanding capability required for practical applications, including code unit perception, intra-code unit understanding, inter-code unit relation understanding, and long documentation understanding.

\item[$\bullet$] \textbf{Extra-long Code Context.} Each task contains around 500 gathered long codes. The lengths of examples change in 0$\sim$8K, 8$\sim$16K, 16$\sim$32K, 32$\sim$64K, and 64$\sim$128K following the normal distribution, which far exceeds the maximum length of 36.5K supported by existing benchmarks \cite{an2023eval}.

\item[$\bullet$] \textbf{Real-world Repository.} The benchmark is collected from real-world code repositories. Long code consists of one or more real code file contents, instead of being composed of multiple independent code units like current benchmarks \cite{liu2024repoqa}.

\item[$\bullet$] \textbf{Reducing Data Contamination.} We collect up-to-date code repositories that are created after 2024-06 from GitHub\footnote{https://github.com/}, which are later than most prevailing LCLMs' cut-off dates thus reducing the risk of data contamination.

\end{itemize}

We evaluate 9 popular LCLMs, which contain 6 general models (\ie GPT-4o \cite{GPT-4o}, Claude-3.5-Sonnet \cite{claude-3.5}, Gemini-1.5-Flash \cite{team2024gemini}, DeepSeek-V2.5 \cite{bai2023longbench}, Mistral-v0.3 \cite{jiang2023mistral}, and Phi-3.5 \cite{abdin2024phi} ) and 3 code models (\ie DeepSeek-Coder-V2 \cite{bai2023longbench},  Qwen2.5-Coder \cite{hui2024qwen2}, CodeLlama \cite{roziere2023code}) on \method. 
The experimental results reveal key limitations in current LCLMs' capabilities for long code understanding. Especially, \textbf{LCLMs’ performance drops dramatically when the long code length is greater than 32K, falling far short of their claimed context windows such as 128K-1M tokens}. 
In the four aspects, \textbf{inter-code unit relation understanding is the most challenging for LCLMs.}
Our findings provide valuable insights for optimizing LCLMs and driving advancements in software engineering.

\begin{table*}[htbp]
  \centering
\setlength{\abovecaptionskip}{0.1cm}
\setlength{\belowcaptionskip}{-5mm}
\caption{The comparison between existing benchmarks and \method. \#Num is the abbreviation of number. \#Div Tasks refers to diverse tasks. \#High Disp represents high dispersion.  \#Max-L and \#Avg-L mean the maximum length and the average length of long code. \#Trunk-L means whether each example has the length label. \#Doc refers to documentation related to repositories. \#Task represents the number of tasks (\ie examples).}
  \resizebox{\linewidth}{!}{
    \begin{tabular}{l|ccc|ccc|ccc|c|c}
    \toprule
    \multirow{2}[2]{*}{Benchmark} & \multicolumn{3}{c|}{Comprehensive Code Tasks}              & \multicolumn{3}{c|}{Extra-long Data} & \multicolumn{3}{c|}{Real-world Repository} & \multicolumn{1}{c|}{Reduce Data Leaking} & \multicolumn{1}{c}{Data Scale} \\
          & \multicolumn{1}{c}{\#Num} &  \multicolumn{1}{c}{\#Div Tasks}  & \multicolumn{1}{c|}{\#High Disp} & \multicolumn{1}{c}{\#Max-L} & \multicolumn{1}{c}{\#Avg-L} & \multicolumn{1}{c|}{\#Length-L} & \multicolumn{1}{c}{Code} & \multicolumn{1}{c}{\#Doc}  & \multicolumn{1}{c|}{\#Num} & \multicolumn{1}{c|}{Data Time}  & \multicolumn{1}{c}{\#Task} \\
    \midrule
    \multicolumn{12}{l}{\textit{\textbf{The second category benchmarks (Only some benchmarks are listed)}}} \\
    \midrule
    LongBench~[\citenum{bai2023longbench}]    &  2 &  \ding{55} &  \ding{55}  &  --  & 0.4K &  \ding{55}  &  Function &  \ding{55} &  -- &  2023.02--2023.08     &  1,000  \\
    LC-Arena~[\citenum{bogomolov2024long}]   &  6 &  \ding{55} &  \ding{55}  & -- & -- & \ding{55}   & File  & \ding{55}  & 62  &  2023.01--2024.05  & --   \\
    LONGPROC~[\citenum{ye2025longproc}]   & 1  & \ding{55}  &  \ding{51}  & -- & 2K & \ding{51}  & Function  &  \ding{55} &  0   & No Limit  & 200  \\
    DevEval~[\citenum{li2024deveval}]   & 1  & \ding{55}  &  \ding{51}  & -- & 0.3K & \ding{51}  & File  &  \ding{55} &  164   & 2023.11-2024.02  & 1,825  \\
    \midrule
    \multicolumn{12}{l}{\textit{\textbf{The first category benchmarks}}} \\
    \midrule
    RepoQA~[\citenum{liu2024repoqa}] &    1   &  \ding{55}    &   \ding{55}   & 16K  &  --   &  \ding{55}   &   Function    &  \ding{55}   &  50     &   No Limit    &  500    \\ 
    L-Eval~[\citenum{an2023eval}] &    1   &  \ding{55}    &   \ding{55}   & 36.5K  &  31.5K   &  \ding{55}   &   Function    &  \ding{55}   &  0    &   No Limit    &  90    \\ 
    \rowcolor{blue!20} \method   & 8 & \ding{51} & \ding{51}  & 128K & 54.8K & \ding{51}    & File & \ding{51}   & 116  & 2024.06--2024.11   &  3,983  \\
    \bottomrule
    \end{tabular}%
        }
    \label{tab:benchmark comparision}%
\end{table*}%

\section{Related Works} \label{related works}
\paragraph{Benchmarks on Long Code Understanding.} 

Existing studies can be mainly categorized into two types. 
The first category predominantly explores the long code understanding ability required in downstream applications like the needle-in-a-haystack task, but they face significant limitations. RepoQA \cite{liu2024repoqa} is a pioneer introduced needle function retrieval task. However, the single task is insufficient to evaluate the complex long code understanding ability. L-Eval \cite{an2023eval}, a widely-used benchmark, has limitations as well. It constructs artificial "long code" by concatenating independent code snippets, ignoring context dependency in real-world source code. 
These limitations highlight the need for a more comprehensive and reliable benchmarking approach for evaluating LCLMs' long code understanding ability. 
The second category includes research like LCArea \cite{bogomolov2024long}. They verify LCLMs' long code understanding ability by measuring their performance on downstream tasks such as LongBench \cite{bai2023longbench}, SWE-bench \cite{jimenez2023swe}, and EvoCodeBench \cite{li2024evocodebench}. Although they provide an intuitive way to assess LCLMs, it suffers from a significant drawback. The performance of downstream tasks is confounded by multiple factors, and these works do not decouple the long code understanding ability independently, which 
is orthogonal to our objective-evaluating LCLMs' capacity to understand and reason about real-world, dependency-rich, long code.

\paragraph{Long Context Language Models.}

Recent studies have explored diverse ways to extend large language models' context window size. The direct way is to fine-tune models on long sequences \cite{wu2021recursively}, but they are often effort-costing.
Some approaches involve additional fine-tuning for a longer context \cite{xiong2023effective, chen2023clex, chen2023extending, chen2023fortify, peng2023yarn}. They down-scale the input position indices to match the original context window size of models with several training steps. 
There are also some training-free studies, which use window attention to clip the long sequences \cite{han2023lm, ding2023longnet, xiao2023efficient}. 
Concurrently, a series of approaches modify the relative distance to extend the extrapolation length \cite{zhang2024hirope, jin2024llm}. In this paper, we construct a comprehensive benchmark to evaluate LCLMs' long code understanding ability.

\vspace{-2mm}

\section{\textbfmethod Benchmark}
\label{benchmark}

\begin{figure}[t]
\centering
\setlength{\belowcaptionskip}{-5mm}
\setlength{\abovecaptionskip}{0.1cm}
\includegraphics[width=\columnwidth]{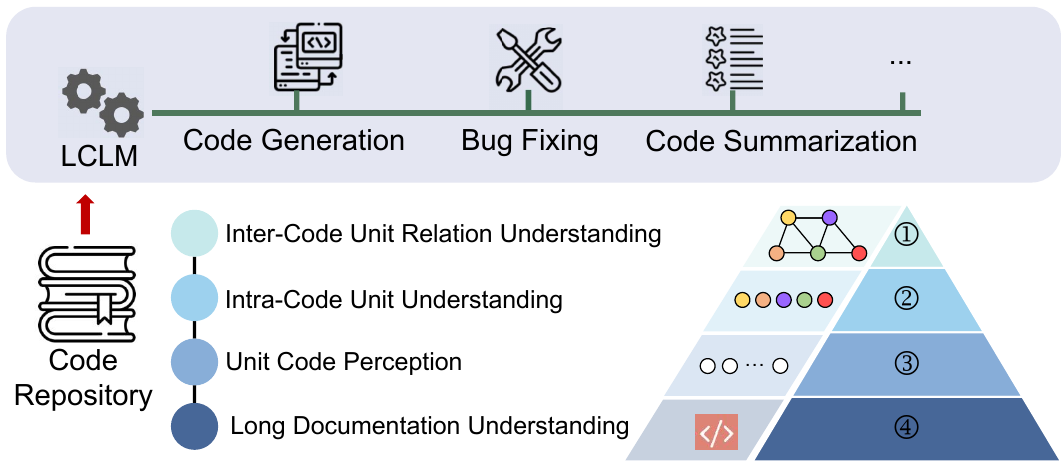}
\caption{Four understanding aspects in \method.}
\label{figure: jzt}
\end{figure}

In this section, we first introduce long code understanding tasks (\textsection \ref{tasks}). Then, we describe the construction process of \method (\textsection \ref{benchmark construction}). Finally, we present the evaluation metrics (\textsection \ref{subsec:auto_eval}).

\subsection{Tasks} \label{tasks}

In real-world software development, long code understanding is usually oriented towards code repositories. These repositories take functions as basic code units, establish relations among units to achieve complex applications, and introduce documentation to describe code-related information. LCLMs with good long code understanding abilities should be able to perceive and understand basic code units, relations among units, and associated code documentation, which is essentially required for dealing with downstream tasks such as repository-level code generation, issue resolving, and long code summarization. In this paper, we propose \method to comprehensively evaluate LCLMs' long code understanding ability from four aspects: code unit perception, intra-code unit understanding, inter-code unit relation understanding, and long documentation understanding.
Four aspects of \method are shown in Figure \ref{figure: jzt}

These long code understanding tasks share the following procedure: given an instruction, long code, and anchor input, LCLMs output the desired answer. The instruction briefly describes the request of each task. The anchor input is the detailed demand such as a code unit. In \method, instruction and anchor input are generally short, while long code is long containing 0$\sim$128K tokens.

\subsubsection{Code Unit Perception}

Understanding long code, particularly in code repositories, requires identifying its numerous functions, as they form the foundational code unit for comprehending long code's overall functionality. 
In this paper, we treat a function as the code unit and introduce the code unit perception task to evaluate LCLMs' code unit identifying ability in long code. 
Concretely, this task requires LCLMs to identify all defined functions in long code and return their corresponding function names, where long code is composed of one or more code file contents collected from real-world repositories. 
This ability is the cornerstone for downstream tasks.

\subsubsection{Intra-Code Unit Understanding}

Based on code unit perception, we further evaluate LCLMs' ability to understand the internal logic and semantics of code units. 

\paragraph{Code Unit Data Flow Analysis.}
In this section, we propose the code unit data flow analysis task that verifies whether LCLMs can understand the internal logic of code units by tracking data flow to a certain extent. Given a code unit in long code and a variable name in the code unit, LCLMs are required to figure out lines where the value of the given variable changes by tracing data flow. For example, after executing the line ``upper\_string += 2", the value of variable ``upper\_string" changes. The given code unit is randomly selected from long code and its position in long code is also random. Evaluating code unit understanding ability is significant for LCLMs to discover potential vulnerabilities, repair vulnerabilities, and optimize code.

\paragraph{Code Unit Semantic Analysis.}
In addition to analyzing the data flow of code units, we propose the code unit semantic analysis task to further verify LCLMs' intra-code unit understanding ability. 
The task asks LCLMs to return a code unit from the long code that satisfies the given description. The long code contains real-world code files, where all descriptions of code units are removed to prevent LCLMs from acquiring clues in them. 
This task requires LCLMs understand the semantics of code units and then return the desired code unit.

\subsubsection{Inter-Code Unit Relation Understanding}

In real-world long code, especially in code repositories, code units are non-standalone. Grasping code unit relations is essential for LCLMs to understand the complex functionality of long code, where code unit relations mainly containing dependency relations and semantic relations. The dependency relation indicates the calls among code units. The semantic relation focuses on the functional similarities of code units.

\paragraph{Dependency Relation Analysis.}
\textbf{(T1)}
Given a code unit, this task requires LCLMs to find code units that are invoked by the given unit from long code, where the long code covers one or multiple code files and is collected from the same repository with the given code unit. The dependency relation analysis ability can assist LCLMs correctly identifying other code units related to vulnerable units and determining vulnerability scopes in real-world applications.
\textbf{(T2)}
Considering that in real-world applications, apart from long code as LCLMs' input, developers usually use natural language requirements to interact with LCLMs. Thus, LCLMs also need to understand dependency relation between long code and requirements. Given a natural language description, this task requires LCLMs to find code units from the long code that are invoked for generating the desired code that satisfies the given description.
The ability can ensure that LCLMs sucessfully invoke existing code units in repository-level code generation and correctly integrate generated code into the current repository.

\paragraph{Semantic Relation Extraction.}
Even if two code units have no dependency relationship, they might be semantically similar such as having similar implementation or logic. This section analyzes semantic relations of code units in long code. 
\textbf{(T1)} 
Given a code unit, this task asks LCLMs to extract semantically similar code units with the given unit from long code. Extracting semantic relations of units can effectively help LCLMs to improve software development efficiency by reusing similar code units when programming, and enhance software maintainability such as finding potentially similar vulnerabilities among semantically similar units.
\textbf{(T2)}
As described in dependency relation analysis (T2), understanding the semantical relations of code units and natural language requirements is more in line with practical development scenarios. For instance, in repository-level code generation, developers input a requirement to LCLMs. LCLMs can find semantically similar units with the given requirement from the current repository and then refer to these units to generate desired codes. 
In this task, LCLMs need to extract semantically similar code units to a given requirement, which challenges LCLMs to understand the semantics of units in long code and reason their semantic relations.

\subsubsection{Long Documentation Understanding}
In real-world software engineering, code documentation also plays a crucial role. It encompasses a diverse range of code-related information including descriptions of code units, usage patterns, architecture designs, and more. Consequently, it is essential not merely to verify the long code understanding ability, but also to analyze the long code documentation understanding ability of LCLMs. We introduce the long documentation understanding task. Given long documentation and a code unit name such as a function name contained in the documentation, this task requires LCLMs to extract the related information to the unit name. We ensure that the long documentation contains the related knowledge of the given unit name, aiming to effectively evaluate LCLMs. Analyzing long documentation is critical to verify the performance of LCLMs in real-world software development.

\begin{table}[t!]
  \centering
  \setlength{\abovecaptionskip}{0.1cm}
  \caption{Statistics of \method. \#Num means the number of examples in each task. \#C-File represents whether the output can be obtained by aggregating cross-file content. \#Avg-L is the average length of the output. \#Gran means the granularity of the output.}
  \resizebox{1\columnwidth}{!}{
    \begin{tabular}{lcc|ccc}
    \toprule
    \multirow{2}[4]{*}{Task} & \multicolumn{2}{c|}{Input}   & \multicolumn{3}{c}{Output} \\
\cmidrule{2-6}     & \multicolumn{1}{c}{\#Num} & \multicolumn{1}{c|}{Format} & \multicolumn{1}{c}{\#C-File} & \multicolumn{1}{c}{\#Avg-L} & \multicolumn{1}{c}{\#Gran} \\
    \midrule
    CU\_P     &  487 & Code      &  \ding{55} & 0.4K  & Name  \\
    CU\_SA    & 500 &  Code  & \ding{55}  & 0.2K & Function \\
    \midrule
    CU\_DFA      & 500   &  Code  & \ding{55}   & 0.03K & Line \\
    \midrule
    DRA\_T1     & 500   &  Code & \ding{51}  &  0.3K  & Function \\
    DRA\_T2    &500    & Code  & \ding{51}  & 0.3K  & Function \\
    SRE\_T1      & 500   &  Code  & \ding{51}  & 1.0K      & Function \\
    SRE\_T2   &  500  & Code & \ding{51}  & 1.1K  & Function \\
    \midrule
    LDU    & 500  &  Document  & \ding{55}  & 0.7K  & NL \\
    \bottomrule
    \end{tabular}%
     }
     \vspace{-0.4cm}
     \label{tab:Statistics of benchmark}%
\end{table}%

\vspace{-1mm}

\subsection{Benchmark Construction}  \label{benchmark construction}

The pipeline for constructing \method encompasses 6 stages as follows.

\noindent \textbf{Stage \ding{182}: Repository Selection.}
Referring to the TIOBE index \cite{Tiobe} for programming language popularity, the most popular language is Python. Thus, we conduct experiments on Python and will expand to other languages in the future. In Python source platforms, PyPI is a rich Python package index tool. We identify the top 10 popular programming topics in PyPI and obtain the top 50 packages with the most stars in each topic. We then select repositories following four criteria: open-source repositories, created after 2024-06, non-fork and non-malicious, and more than 50 stars. Finally, we crawl these repositories from GitHub and obtain 116 real-world repositories.

\noindent \textbf{Stage \ding{183}: Code Parse.}
We use \textit{tree-sitter}\footnote{https://tree-sitter.github.io/tree-sitter/} to design a code parser. This parser identifies code units defined in repositories, traces the definition of code units, and analyzes unit relations: First, it performs static analysis of each file in a repository and extracts unit names defined in it. Then, it executes unit symbol navigation, finding the definitions of all code units in the repository. Finally, it extracts unit names invoked in a code unit to grasp their dependency relations. Combining the three steps, the parser can traverse predefined code units within a repository and obtain intricate dependencies. To obtain semantic relations among units, we apply an advanced embedding model to encode code units, and use the cosine similarities of their representations to measure semantic relations.

\noindent \textbf{Stage \ding{184}: Requirement Collection.}
We extract requirements contained in the signature of code units with \textit{tree-sitter} and invite two developers to check requirements. Not that it is enough to construct examples for T2-type tasks even if some code units do not contain requirements in repositories.

\noindent \textbf{Stage \ding{185}: Documentation Annotation.}
We collect documentation from collected code repositories and select the documentation with standard: being easy to distinguish the related information of code units. Then we invite two developers to manually label 500 examples.

\noindent \textbf{Stage \ding{186}: Deduplication.}
For tasks whose input is code unit, we exclude trivial units (\eg empty or initialization functions). To ensure the quality of \method, we randomly select files from repositories as long codes, and ensure long codes of all examples in each task are non-repetitive.

\noindent \textbf{Stage \ding{187}: Benchmark Construction.}
Based on the above stages, we construct around 500 examples for each task supporting the maximum length of 128K tokens.
We make \method satisfy following goals: comprehensive tasks from practical applications, extra-long code context, real-world repositories, and reducing data contamination.

\subsection{Automatic Evaluation} \label{subsec:auto_eval}

We focus on evaluating the long code understanding ability required for LCLMs to complete downstream tasks. Tasks in \method commonly require LCLMs to retrieve dispersive snippets from long code and execute reasoning. Thus, we mainly adopt metrics used in retrieval tasks to measure LCLMs, including recall and precision.

For proper evaluation, we refine existing metrics according to the output granularity of tasks: 
\ding{182} \textit{Output with code lines} refers to the code unit data flow analysis task, which extracts several lines from long code. We first use exact match (\textbf{EM}) to measure each line in outputs. Then, we calculate recall and precision based on EM, acquiring \textbf{EM-R} and \textbf{EM-P} metrics.
\ding{183} \textit{Output with code unit names} is related to the code unit perception task. We first find the correct names in output and then calculate the longest common subsequence (LCS) of unit names between output and ground truth, since it not only reflects whether the unit name is correct but also indicates LCLMs' ability to perceive the position of units in long code. Then, we calculate recall and precision based on LCS, named \textbf{LCS-R} and \textbf{LCS-P}.
\ding{184} \textit{Output with code units} refers to code unit semantic analysis and inter-code unit relation understanding tasks. Code unit semantic analysis only returns one unit. We use \textbf{CodeBLEU} as its metric, where it is a popular metric to indicate the consistency of two code sequences. For inter-code unit relation understanding, it returns multiple code units. We first determine whether the name of each generated unit is in ground truth. Then, we calculate CodeBLEU of each unit if it exists in the ground truth. For the code unit whose name is not in the ground truth, its CodeBLEU value is set to 0. Based on CodeBLEU value of each unit, we finally calculate CodeBLEU-based recall and precision, dubbed \textbf{CB-R} and \textbf{CB-P}.
\ding{185} \textit{Output with descriptions} is related to the long documentation understanding task. Considering that documentation contains many text descriptions, we employ \textbf{BLEU} for evaluation, which is commonly used to measure the consistency between two sequences in the natural language process field.

These automatic evaluations can effectively measure LCLMs' long code understanding ability since the outputs of each task are deterministic and task' features are incorporated into these metrics. Figure \ref{figure: Reliable Evaluation Metrics} shows the automatic evaluation correlates well with human annotation, which further demonstrates the reliability of our automatic metrics.

\begin{table*}[htbp]
  \centering
  \setlength{\abovecaptionskip}{0.1cm}
  \setlength{\belowcaptionskip}{-5mm}
  \caption{The performance of LCLMs on \method. 
  We only report recall-based results (EM-R, LCS-R, and CB-R) due to page limitation. The precision-based results (EM-P, LCS-P, and CB-P) can be found in Appendix \ref{sec:appendix-rq2}.} 
    \label{tab: main results}
    \resizebox{1\linewidth}{!}{
    \begin{tabular}{lcc|c|cc|cccc|c|c}
    \toprule
    \multirow{2}[2]{*}{} & \multicolumn{1}{c}{\multirow{2}[2]{*}{\textbf{\#Param}}} & \multicolumn{1}{c|}{\multirow{2}[2]{*}{\textbf{Context Size}}} & \multicolumn{9}{c}{\textbf{Task}}  \\
          &       &       & \multicolumn{1}{c|}{\textbf{CU\_P}} & \multicolumn{1}{c}{\textbf{CU\_SA}} & \multicolumn{1}{c|}{\textbf{CU\_DFA}} & \multicolumn{1}{c}{\textbf{DRA\_T1}} & \multicolumn{1}{c}{\textbf{DRA\_T2}} & \multicolumn{1}{c}{\textbf{SRE\_T1}} & \multicolumn{1}{c|}{\textbf{SRE\_T2}} & \multicolumn{1}{c|}{\textbf{LDU}}  & \multicolumn{1}{c|}{\textbf{\#Avg}}   \\
    \midrule
    \textit{Code Models} & \multicolumn{10}{c}{\textit{Open-Source LCLMs}} \\
    Qwen2.5-Coder     & 7.6B & 128K  & \cellcolor{yellow!30}43.47 & \cellcolor{green!30}71.06 & \cellcolor{green!30}74.01 & \cellcolor{blue!25}30.38  & \cellcolor{blue!25}9.59 & \cellcolor{blue!25}24.34  & \cellcolor{blue!25}21.81   & \cellcolor{red!30}21.83   & 37.06 \\
    DeepSeek-Coder-V2     & 15.7B &  128K  & \cellcolor{yellow!30}38.67  & \cellcolor{green!30}65.21 & \cellcolor{green!30}48.42 & \cellcolor{blue!25}47.26 & \cellcolor{blue!25}22.92 &  \cellcolor{blue!25}24.61     & \cellcolor{blue!25}26.21 & \cellcolor{red!30}50.69     & 40.49 \\
    CodeLlama    &  33.7B  & 16K  & \cellcolor{yellow!30}68.57  & \cellcolor{green!30}62.41  &\cellcolor{green!30}79.87  & \cellcolor{blue!25}68.82  &\cellcolor{blue!25} 34.94 & \cellcolor{blue!25}44.48 &  \cellcolor{blue!25}36.34  &\cellcolor{red!30} 46.92  & 55.29 \\
    \midrule
    \textit{General Models} & \multicolumn{10}{c}{\textit{Open-Source LCLMs}}  \\
    Phi-3.5     &  3.8B  &128K  & \cellcolor{yellow!30}39.92  & \cellcolor{green!30}46.75  & \cellcolor{green!30}49.52 &\cellcolor{blue!25}30.76  & \cellcolor{blue!25}9.66 & \cellcolor{blue!25}18.99  & \cellcolor{blue!25}14.48    & \cellcolor{red!30}34.14    & 30.53  \\
    Mistral-v0.3     &  7.3B  & 32K & \cellcolor{yellow!30}57.42  & \cellcolor{green!30}63.90 & \cellcolor{green!30}58.00  & \cellcolor{blue!25}46.66 & \cellcolor{blue!25}18.92 & \cellcolor{blue!25}33.91  & \cellcolor{blue!25}32.50 & \cellcolor{red!30}58.64    & 46.24 \\
    DeepSeek-V2.5     &  236B  &  128K   & \cellcolor{yellow!30}\textbf{70.58} & \cellcolor{green!30}82.11  & \cellcolor{green!30}77.47 & \cellcolor{blue!25}72.25 & \cellcolor{blue!25}\textbf{56.80}  & \cellcolor{blue!25}\textbf{49.08} & \cellcolor{blue!25}\textbf{47.42}  & \cellcolor{red!30}85.85    & \textbf{67.70} \\
\cmidrule{2-12}          & \multicolumn{11}{c}{\textit{Proprietary Source LCLMs}} \\
    Claude-3.5-Sonnet    & --  & 200K & \cellcolor{yellow!30}43.82 & \cellcolor{green!30}40.60 & \cellcolor{green!30}45.65 & \cellcolor{blue!25}29.37 & \cellcolor{blue!25}28.70 & \cellcolor{blue!25}26.55 & \cellcolor{blue!25}27.77 & \cellcolor{red!30}41.81    &35.53  \\
    Gemini-1.5-Flash     &  --  & 1000K & \cellcolor{yellow!30}58.45  & \cellcolor{green!30}83.46  & \cellcolor{green!30}80.37 & \cellcolor{blue!25}\textbf{72.51} & \cellcolor{blue!25}46.42 & \cellcolor{blue!25}39.84 & \cellcolor{blue!25}38.69  &\cellcolor{red!30} 81.43    &61.39 \\
    GPT-4o     & --  & 128K  & \cellcolor{yellow!30}56.42 & \cellcolor{green!30}\textbf{86.76}  & \cellcolor{green!30}\textbf{87.87} & \cellcolor{blue!25}71.58 &\cellcolor{blue!25} 48.88 & \cellcolor{blue!25}44.45  & \cellcolor{blue!25}43.14  & \cellcolor{red!30}\textbf{87.54}    & \textbf{65.83} \\
    \bottomrule
    \end{tabular}%
     }
\end{table*}%

\vspace{-2mm}

\section{Experiment Setups} \label{sec: evaluation}

In this section, we aim to answer the following research questions through a series of experiments.

\noindent\textbf{RQ1. How is the long code understanding ability of LCLMs?} We evaluate LCLMs' long code understanding ability on \method in \S\ref{Long Code Understanding Capability}.

\noindent\textbf{RQ2. What is performance of LCLMs across long code lengths?} We explore the performance of LCLMs on long code with different lengths. \S\ref{Performance across Long Code Lengths} demonstrates LCLMs' performance comparison across long code lengths.

\noindent\textbf{RQ3. How do developers select models in real-world application scenarios?} Based on the experimental results, we summarize the empirical lessons we learned, aiming to help developers select suitable LCLMs ~(\S\ref{subsec:downstream_task}).

\subsection{Base LCLMs}
We evaluate 9 advanced LCLMs on \method, which contain 6 general models (\ie GPT-4o \cite{GPT-4o}, Claude-3.5-Sonnet \cite{claude-3.5}, Gemini-1.5-Flash \cite{team2024gemini}, DeepSeek-V2.5 \cite{bai2023longbench}, Mistral-v0.3 \cite{jiang2023mistral}, and Phi-3.5 \cite{abdin2024phi} ) and 3 code models (\ie DeepSeek-Coder-V2 \cite{bai2023longbench},  Qwen2.5-Coder \cite{hui2024qwen2}, CodeLlama \cite{roziere2023code}). DeepSeek-R1 and GPT-o3-mini are newly released LCLMs, but their availability through invoking API is unstable or limited to usage frequency. We can not present their performance now and will evaluate them in the future.

\subsection{Experimental Setup}
We use greedy search for all experiments. We evaluate LCLMs within their maximum context window length. In the semantic relation extraction task, we use an advanced embedding model stella\_en\_400M\_v5 to encode code units and natural language requirements with the 1,024 dimension version, and select the top five similar code units by calculating the cosine similarities of their embeddings. Constrained by computing resources, we evaluate DeepSeek-V2.5 by invoking API\footnote{https://api-docs.deepseek.com/api/deepseek-api} provided by DeepSeek. 
Although the context length of DeepSeek-V2.5 achieves 128K tokens, the API provided by DeepSeek only supports up to 60K tokens. Thus, we analyze DeepSeek-V2.5 on long codes with 0$\sim$64K tokens.

\section{Results and Analysis} \label{sec: evaluation}

\begin{figure*}[t]
\centering
\setlength{\abovecaptionskip}{0.1cm}
\setlength{\belowcaptionskip}{-5mm}
\includegraphics[width=\linewidth]{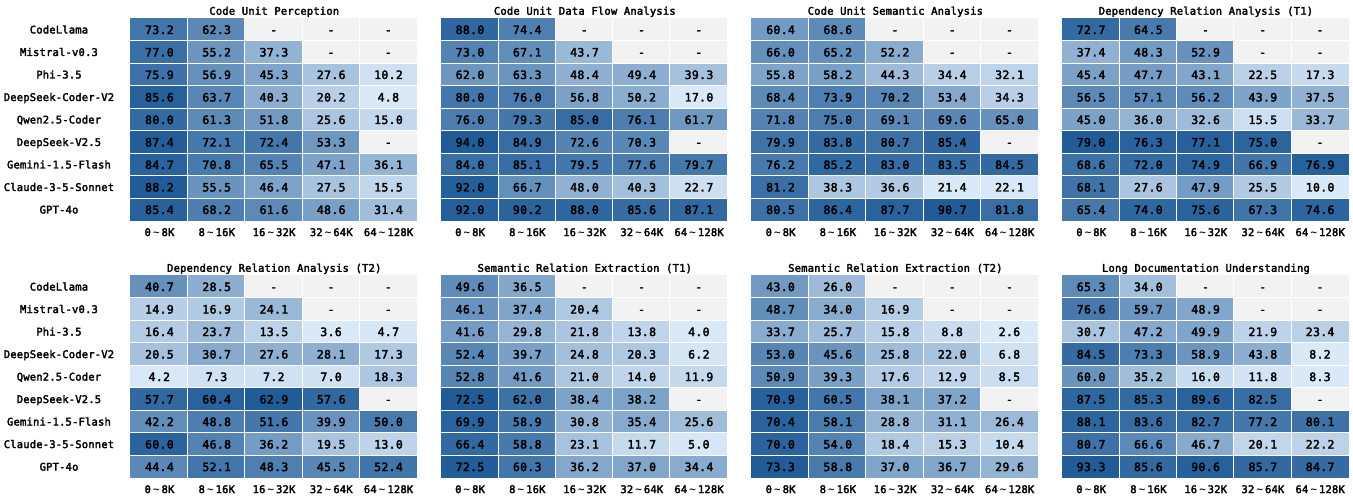}
\caption{Performance comparison across tasks and long code lengths on \method (grey blocks indicate unavailable configurations). The rate of performance degradation exhibits task-specific and model-specific patterns. }
\label{Figure: heat map}
\end{figure*}

\subsection{Long Code Understanding Capability} \label{Long Code Understanding Capability}

Table~\ref{tab: main results} presents the performance of LCLMs on \method.  
For LCLMs with a context length of less than 128K tokens, we only evaluate them within their supporting maximum lengths.

\paragraph{Comparison across LCLMs.}
We observe significant performance gaps among LCLMs. 
For similar-scale models, the performance of code LCLMs is better than the counterpart of general models on most tasks. For example, Qwen2.5-Coder outperforms Phi-3.5 24.31\% on intra-code unit understanding (the second-aspect task) in terms of CodeBLEU, and DeepSeek-Coder-V2 achieves 11.75\% average improvements on inter-code unit relation understanding (the third-aspect task) in CB-R. Among open-source models, DeepSeek-V2.5 performs the best, which is related to its large number of parameters. In proprietary source models, GPT-4o achieves the best performance, while Claude-3.5-Sonnet is not satisfactory. 

\paragraph{Comparison across Tasks.}
In the four aspects, LCLMs perform the best in code unit perception and long documentation understanding, and achieve moderate results in intra-unit code understanding. Inter-code unit relation understanding is the most challenging. 
Their performances are reasonable. Because it's a fundamental ability for LCLMs to understand code documentation and perceive code units. Based on this, LCLMs understand intra-unit code and then analyze inter-code unit relations, thereby comprehending long code.
When observing the third-aspect task (\ie inter-code unit relation understanding) closely, it can be seen that the performance of dependency relation analysis is lower than that of semantic relation extraction. 
We also find that no LCLM outperforms others on all tasks. For instance, in terms of scale-similar code models and general models, the former outperforms the latter in code unit perception, while code models perform worse than general models in long documentation understanding.

\subsection{Performance across Long Code Lengths} \label{Performance across Long Code Lengths}
To address RQ2, we classify examples into five buckets according to the long code length, including 0$\sim$8K, 8$\sim$16K, 16$\sim$32K, 32$\sim$64K, and 64$\sim$128K on each task. We conduct experiments on these classes to investigate the true context ability supported by LCLMs in long code understanding. Figure \ref{Figure: heat map} and Appendix \ref{sec:appendix-rq2} shows the results of LCLMs across long code lengths. 

Our experimental results reveal key limitations in current LCLMs’ capabilities for long code understanding. 
We can also observe a negative correlation between the long code length and the performance of LCLMs. 
LCLMs' performance drops dramatically when the long code length is greater than 32K tokens, falling well short of their claimed 128K$\sim$1M context windows.
Besides, when the long code contains 64$\sim$128K tokens, the performance of LCLMs is near to 10 or even close to 0 on some tasks such as dependency relation analysis and semantic relation extraction.
In addition, the degradation slopes of performances vary by task. For example, there is a large slope in long documentation understanding, which means that LCLMs are suitable for processing relatively short documentation. The slope on code unit understanding including code unit data flow analysis and code unit semantic analysis tasks is relatively small.  LCLMs fail to model code context effectively in their claimed context windows.

\subsection{Empirical Lessons.} \label{subsec:downstream_task}
Based on the above experiments, we summarize the empirical lessons we learned as:
\ding{182} The small-size LCLMs such as Qwen2.5-Coder can satisfy developers' need for code unit perception and understanding if long code length is less than 16K. Otherwise, we suggest choosing larger-size models. 
\ding{183} For understanding long code documentation with more than 32K tokens, it is recommended to use GPT-4o and Gemini-1.5-Flash. Otherwise, Mistral-v0.3 and DeepSeek-Coder-V2 can achieve satisfying performances.
\ding{184} 
For tasks with high requirements for understanding inter-code unit relations, we suggest selecting the most powerful LCLMs that developers can access. Because when long code length exceeds 64K which is common in repository-level tasks, strong LCLMs also achieve weak perfrmances. These poor performances explain why powerful LCLMs perform unsatisfactorily in repository-level downstream tasks. For example, GPT-4o only achieves 4.00\% Success@1 on repository-level code translation benchmark-RepoTransBench \cite{DBLP:journals/corr/abs-2412-17744}.


\section{Discussion}

\subsection{Case Study}

We analyze the outputs of LCLMs, particularly GPT-4o, on the inter-code unit relation understanding task. Appendix \ref{sec:appendix-case} presents two notable examples. GPT-4o often extracts code units that are structurally similar or share overlapping tokens but have distinct or even opposite functionalities. This highlights the need to enhance models' ability to distinguish confusing code units.

\subsection{Code Understanding or Memorization?}

We collect a small number of early-released code repositories from GitHub that LCLMs have potentially encountered during LCLMs training, aiming to analyze the dependency degree of LCLMs on memorization and long code understanding ability. Concretely, we enter only an instruction and anchor input to models, withholding the long code context, and assess their performance. We select tasks where models can work normally even without long code context, such as code unit semantic analysis and dependency relation analysis (T2). As shown in Figure~\ref{figure: memory}, the memorization performance (w/o context) is much lower than the results with long code context, even though models might have met these contexts when training. The $\delta$ score (the results with context minus the performance without context in yellow) relieves the memory phenomenon \cite{yu2023kola} and also reveals the significant importance for measuring LCLMs' long code understanding ability.

\subsection{Reliable Evaluation Metrics?}

Reliable evaluation metrics are essential for assessing long code understanding. We measure the consistency between our metrics and human evaluation by selecting 20 GPT-4o outputs on each task and inviting two advanced developers to manually evaluate them. Using Kendall-Tau $\tau$ \cite{kendall1938new}, we found that the average $\tau$ value is at least 0.75 across all tasks, with the minimum value exceeding 0.7 (Figure \ref{figure: Reliable Evaluation Metrics}). This demonstrates a strong correlation between our metrics and human evaluation, confirming the reliability of our metrics.

\vspace{-1mm}

\section{Conclusion}

In this paper, we propose a comprehensive long code understanding benchmark \method. It introduces four aspects (8 tasks) to evaluate LCLMs’ long code understanding ability required for practical applications, including code unit perception, intra-code unit understanding, inter-code unit relation understanding, and long code documentation understanding. Our experimental results reveal key limitations in current LCLMs’ capabilities for long code understanding. 
When the long code length contains more than 32K tokens, 
the performance of LCLMs drops dramatically, falling far short of their claimed 128K$\sim$1M context windows. 
We hope our findings can provide valuable insights for optimizing LCLMs and driving advancements in software engineering.

\section*{Limitations}

This paper proposes a benchmark - \method to evaluate the long code understanding ability of long context language models from four aspects which are essential capabilities required for LCLMs to complete real-world downstream tasks. 
Based on \method, we evaluate 9 popular LCLMs and analyze their strengths and shortcomings. We think that DevEval has three limitations.

\ding{182} \method is a monolingual benchmark (\ie requirements in English and code in Python) and ignores other languages. In practice, LLMs require understanding requirements in different natural languages (\eg Chinese, Spanish) and generating programs in various programming languages (\eg Java, C). Thus, we plan to build a multilingual
\method in future work.

\ding{183} Most recently, there are a few newly released LCLMs such as DeepSeek-R1 and OpenAI o3-mini-high. Constrained by the availability or stabilization of API, we do not provide the performance of these newly released models. In the future, we will evaluate their long code understanding abilities on \method.

\ding{184} In our experiments, we only consider the long code with 0$\sim$128K tokens, although some LCLMs have supported longer context windows, \eg Claude-3-5-Sonnet with 200K context size and even Gemini-1.5-Flash with 1000K context size. We will continue to update and evolve our benchmark, in order to support \method to measure LCLMs on longer codes.


\bibliography{main}

\begin{thebibliography}{33}
\providecommand{\natexlab}[1]{#1}

\bibitem[{cla(2024)}]{claude-3.5}
 2024.
\newblock Claude 3.5 haiku.
\newblock \emph{https://www.anthropic.com/claude/haiku}.

\bibitem[{GPT(2024)}]{GPT-4o}
 2024.
\newblock Gpt-4o.
\newblock \emph{https://openai.com/index/hello-gpt-4o/}.

\bibitem[{Tio(2024)}]{Tiobe}
 2024.
\newblock Tiobe-index.
\newblock \emph{https://www.tiobe.com/tiobe- index/}.

\bibitem[{Abdin et~al.(2024)Abdin, Jacobs, Awan, Aneja, Awadallah, Awadalla, Bach, Bahree, Bakhtiari, Behl et~al.}]{abdin2024phi}
Marah Abdin, Sam~Ade Jacobs, Ammar~Ahmad Awan, Jyoti Aneja, Ahmed Awadallah, Hany Awadalla, Nguyen Bach, Amit Bahree, Arash Bakhtiari, Harkirat Behl, et~al. 2024.
\newblock Phi-3 technical report: A highly capable language model locally on your phone.
\newblock \emph{arXiv preprint arXiv:2404.14219}.

\bibitem[{An et~al.(2023)An, Gong, Zhong, Zhao, Li, Zhang, Kong, and Qiu}]{an2023eval}
Chenxin An, Shansan Gong, Ming Zhong, Xingjian Zhao, Mukai Li, Jun Zhang, Lingpeng Kong, and Xipeng Qiu. 2023.
\newblock L-eval: Instituting standardized evaluation for long context language models.
\newblock \emph{arXiv preprint arXiv:2307.11088}.

\bibitem[{Bai et~al.(2023)Bai, Lv, Zhang, Lyu, Tang, Huang, Du, Liu, Zeng, Hou et~al.}]{bai2023longbench}
Yushi Bai, Xin Lv, Jiajie Zhang, Hongchang Lyu, Jiankai Tang, Zhidian Huang, Zhengxiao Du, Xiao Liu, Aohan Zeng, Lei Hou, et~al. 2023.
\newblock Longbench: A bilingual, multitask benchmark for long context understanding.
\newblock \emph{arXiv preprint arXiv:2308.14508}.

\bibitem[{Bi et~al.(2024)Bi, Wan, Wang, Zhang, Guan, Lu, Zhang, Sui, Jin, and Shi}]{bi2024iterative}
Zhangqian Bi, Yao Wan, Zheng Wang, Hongyu Zhang, Batu Guan, Fangxin Lu, Zili Zhang, Yulei Sui, Hai Jin, and Xuanhua Shi. 2024.
\newblock Iterative refinement of project-level code context for precise code generation with compiler feedback.
\newblock \emph{arXiv preprint arXiv:2403.16792}.

\bibitem[{Bogomolov et~al.(2024)Bogomolov, Eliseeva, Galimzyanov, Glukhov, Shapkin, Tigina, Golubev, Kovrigin, van Deursen, Izadi et~al.}]{bogomolov2024long}
Egor Bogomolov, Aleksandra Eliseeva, Timur Galimzyanov, Evgeniy Glukhov, Anton Shapkin, Maria Tigina, Yaroslav Golubev, Alexander Kovrigin, Arie van Deursen, Maliheh Izadi, et~al. 2024.
\newblock Long code arena: a set of benchmarks for long-context code models.
\newblock \emph{arXiv preprint arXiv:2406.11612}.

\bibitem[{Chen et~al.(2023{\natexlab{a}})Chen, Li, Meng, Liang, and Bing}]{chen2023clex}
Guanzheng Chen, Xin Li, Zaiqiao Meng, Shangsong Liang, and Lidong Bing. 2023{\natexlab{a}}.
\newblock Clex: Continuous length extrapolation for large language models.
\newblock \emph{arXiv preprint arXiv:2310.16450}.

\bibitem[{Chen et~al.(2023{\natexlab{b}})Chen, Wong, Chen, and Tian}]{chen2023extending}
Shouyuan Chen, Sherman Wong, Liangjian Chen, and Yuandong Tian. 2023{\natexlab{b}}.
\newblock Extending context window of large language models via positional interpolation.
\newblock \emph{arXiv preprint arXiv:2306.15595}.

\bibitem[{Chen et~al.(2023{\natexlab{c}})Chen, Lv, Lin, Chen, Wu, Huang, Li, and Yan}]{chen2023fortify}
Yuhan Chen, Ang Lv, Ting-En Lin, Changyu Chen, Yuchuan Wu, Fei Huang, Yongbin Li, and Rui Yan. 2023{\natexlab{c}}.
\newblock Fortify the shortest stave in attention: Enhancing context awareness of large language models for effective tool use.
\newblock \emph{arXiv preprint arXiv:2312.04455}.

\bibitem[{Dhulshette et~al.(2025)Dhulshette, Shah, and Kulkarni}]{dhulshette2025hierarchical}
Nilesh Dhulshette, Sapan Shah, and Vinay Kulkarni. 2025.
\newblock Hierarchical repository-level code summarization for business applications using local llms.
\newblock \emph{arXiv preprint arXiv:2501.07857}.

\bibitem[{Ding et~al.(2023)Ding, Ma, Dong, Zhang, Huang, Wang, Zheng, and Wei}]{ding2023longnet}
Jiayu Ding, Shuming Ma, Li~Dong, Xingxing Zhang, Shaohan Huang, Wenhui Wang, Nanning Zheng, and Furu Wei. 2023.
\newblock Longnet: Scaling transformers to 1,000,000,000 tokens.
\newblock \emph{arXiv preprint arXiv:2307.02486}.

\bibitem[{Han et~al.(2023)Han, Wang, Xiong, Chen, Ji, and Wang}]{han2023lm}
Chi Han, Qifan Wang, Wenhan Xiong, Yu~Chen, Heng Ji, and Sinong Wang. 2023.
\newblock Lm-infinite: Simple on-the-fly length generalization for large language models.
\newblock \emph{arXiv preprint arXiv:2308.16137}.

\bibitem[{Hui et~al.(2024)Hui, Yang, Cui, Yang, Liu, Zhang, Liu, Zhang, Yu, Lu et~al.}]{hui2024qwen2}
Binyuan Hui, Jian Yang, Zeyu Cui, Jiaxi Yang, Dayiheng Liu, Lei Zhang, Tianyu Liu, Jiajun Zhang, Bowen Yu, Keming Lu, et~al. 2024.
\newblock Qwen2. 5-coder technical report.
\newblock \emph{arXiv preprint arXiv:2409.12186}.

\bibitem[{Jiang et~al.(2023)Jiang, Sablayrolles, Mensch, Bamford, Chaplot, Casas, Bressand, Lengyel, Lample, Saulnier et~al.}]{jiang2023mistral}
Albert~Q Jiang, Alexandre Sablayrolles, Arthur Mensch, Chris Bamford, Devendra~Singh Chaplot, Diego de~las Casas, Florian Bressand, Gianna Lengyel, Guillaume Lample, Lucile Saulnier, et~al. 2023.
\newblock Mistral 7b.
\newblock \emph{arXiv preprint arXiv:2310.06825}.

\bibitem[{Jimenez et~al.(2023)Jimenez, Yang, Wettig, Yao, Pei, Press, and Narasimhan}]{jimenez2023swe}
Carlos~E Jimenez, John Yang, Alexander Wettig, Shunyu Yao, Kexin Pei, Ofir Press, and Karthik Narasimhan. 2023.
\newblock Swe-bench: Can language models resolve real-world github issues?
\newblock \emph{arXiv preprint arXiv:2310.06770}.

\bibitem[{Jin et~al.(2024)Jin, Han, Yang, Jiang, Liu, Chang, Chen, and Hu}]{jin2024llm}
Hongye Jin, Xiaotian Han, Jingfeng Yang, Zhimeng Jiang, Zirui Liu, Chia-Yuan Chang, Huiyuan Chen, and Xia Hu. 2024.
\newblock Llm maybe longlm: Self-extend llm context window without tuning.
\newblock \emph{arXiv preprint arXiv:2401.01325}.

\bibitem[{Kendall(1938)}]{kendall1938new}
Maurice~G Kendall. 1938.
\newblock A new measure of rank correlation.
\newblock \emph{Biometrika}, 30(1-2):81--93.

\bibitem[{Li et~al.(2024{\natexlab{a}})Li, Li, Zhang, Zhao, Dong, Jin, Li, Huang, and Li}]{li2024evocodebench}
Jia Li, Ge~Li, Xuanming Zhang, Yunfei Zhao, Yihong Dong, Zhi Jin, Binhua Li, Fei Huang, and Yongbin Li. 2024{\natexlab{a}}.
\newblock \href {http://papers.nips.cc/paper\_files/paper/2024/hash/6a059625a6027aca18302803743abaa2-Abstract-Datasets\_and\_Benchmarks\_Track.html} {Evocodebench: An evolving code generation benchmark with domain-specific evaluations}.
\newblock In \emph{Advances in Neural Information Processing Systems 38: Annual Conference on Neural Information Processing Systems 2024, NeurIPS 2024, Vancouver, BC, Canada, December 10 - 15, 2024}.

\bibitem[{Li et~al.(2024{\natexlab{b}})Li, Li, Zhao, Li, Liu, Zhu, Wang, Liu, Fang, Wang, Ding, Zhang, Zhu, Dong, Jin, Li, Huang, Li, Gu, and Yang}]{li2024deveval}
Jia Li, Ge~Li, Yunfei Zhao, Yongmin Li, Huanyu Liu, Hao Zhu, Lecheng Wang, Kaibo Liu, Zheng Fang, Lanshen Wang, Jiazheng Ding, Xuanming Zhang, Yuqi Zhu, Yihong Dong, Zhi Jin, Binhua Li, Fei Huang, Yongbin Li, Bin Gu, and Mengfei Yang. 2024{\natexlab{b}}.
\newblock \href {https://doi.org/10.18653/V1/2024.FINDINGS-ACL.214} {Deveval: {A} manually-annotated code generation benchmark aligned with real-world code repositories}.
\newblock In \emph{Findings of the Association for Computational Linguistics, {ACL} 2024, Bangkok, Thailand and virtual meeting, August 11-16, 2024}, pages 3603--3614. Association for Computational Linguistics.

\bibitem[{Liu et~al.(2024)Liu, Tian, Daita, Wei, Ding, Wang, Yang, and Zhang}]{liu2024repoqa}
Jiawei Liu, Jia~Le Tian, Vijay Daita, Yuxiang Wei, Yifeng Ding, Yuhan~Katherine Wang, Jun Yang, and Lingming Zhang. 2024.
\newblock Repoqa: Evaluating long context code understanding.
\newblock \emph{arXiv preprint arXiv:2406.06025}.

\bibitem[{Peng et~al.(2023)Peng, Quesnelle, Fan, and Shippole}]{peng2023yarn}
Bowen Peng, Jeffrey Quesnelle, Honglu Fan, and Enrico Shippole. 2023.
\newblock Yarn: Efficient context window extension of large language models.
\newblock \emph{arXiv preprint arXiv:2309.00071}.

\bibitem[{Roziere et~al.(2023)Roziere, Gehring, Gloeckle, Sootla, Gat, Tan, Adi, Liu, Remez, Rapin et~al.}]{roziere2023code}
Baptiste Roziere, Jonas Gehring, Fabian Gloeckle, Sten Sootla, Itai Gat, Xiaoqing~Ellen Tan, Yossi Adi, Jingyu Liu, Tal Remez, J{\'e}r{\'e}my Rapin, et~al. 2023.
\newblock Code llama: Open foundation models for code.
\newblock \emph{arXiv preprint arXiv:2308.12950}.

\bibitem[{Team et~al.(2024)Team, Georgiev, Lei, Burnell, Bai, Gulati, Tanzer, Vincent, Pan, Wang et~al.}]{team2024gemini}
Gemini Team, Petko Georgiev, Ving~Ian Lei, Ryan Burnell, Libin Bai, Anmol Gulati, Garrett Tanzer, Damien Vincent, Zhufeng Pan, Shibo Wang, et~al. 2024.
\newblock Gemini 1.5: Unlocking multimodal understanding across millions of tokens of context.
\newblock \emph{arXiv preprint arXiv:2403.05530}.

\bibitem[{Wang et~al.(2024)Wang, Wang, Wang, Guo, Chen, Grundy, Liu, Ma, Mao, Zhang, and Zheng}]{DBLP:journals/corr/abs-2412-17744}
Yanlin Wang, Yanlin Wang, Suiquan Wang, Daya Guo, Jiachi Chen, John~C. Grundy, Xilin Liu, Yuchi Ma, Mingzhi Mao, Hongyu Zhang, and Zibin Zheng. 2024.
\newblock \href {https://doi.org/10.48550/ARXIV.2412.17744} {Repotransbench: {A} real-world benchmark for repository-level code translation}.
\newblock \emph{CoRR}, abs/2412.17744.

\bibitem[{Wu et~al.(2021)Wu, Ouyang, Ziegler, Stiennon, Lowe, Leike, and Christiano}]{wu2021recursively}
Jeff Wu, Long Ouyang, Daniel~M Ziegler, Nisan Stiennon, Ryan Lowe, Jan Leike, and Paul Christiano. 2021.
\newblock Recursively summarizing books with human feedback.
\newblock \emph{arXiv preprint arXiv:2109.10862}.

\bibitem[{Xiao et~al.(2023)Xiao, Tian, Chen, Han, and Lewis}]{xiao2023efficient}
Guangxuan Xiao, Yuandong Tian, Beidi Chen, Song Han, and Mike Lewis. 2023.
\newblock Efficient streaming language models with attention sinks.
\newblock \emph{arXiv preprint arXiv:2309.17453}.

\bibitem[{Xiong et~al.(2023)Xiong, Liu, Molybog, Zhang, Bhargava, Hou, Martin, Rungta, Sankararaman, Oguz et~al.}]{xiong2023effective}
Wenhan Xiong, Jingyu Liu, Igor Molybog, Hejia Zhang, Prajjwal Bhargava, Rui Hou, Louis Martin, Rashi Rungta, Karthik~Abinav Sankararaman, Barlas Oguz, et~al. 2023.
\newblock Effective long-context scaling of foundation models.
\newblock \emph{arXiv preprint arXiv:2309.16039}.

\bibitem[{Ye et~al.(2025)Ye, Yin, He, Zhang, Yen, Gao, Durrett, and Chen}]{ye2025longproc}
Xi~Ye, Fangcong Yin, Yinghui He, Joie Zhang, Howard Yen, Tianyu Gao, Greg Durrett, and Danqi Chen. 2025.
\newblock Longproc: Benchmarking long-context language models on long procedural generation.
\newblock \emph{arXiv preprint arXiv:2501.05414}.

\bibitem[{Yu et~al.(2023)Yu, Wang, Tu, Cao, Zhang-Li, Lv, Peng, Yao, Zhang, Li et~al.}]{yu2023kola}
Jifan Yu, Xiaozhi Wang, Shangqing Tu, Shulin Cao, Daniel Zhang-Li, Xin Lv, Hao Peng, Zijun Yao, Xiaohan Zhang, Hanming Li, et~al. 2023.
\newblock Kola: Carefully benchmarking world knowledge of large language models.
\newblock \emph{arXiv preprint arXiv:2306.09296}.

\bibitem[{Zhang et~al.(2024{\natexlab{a}})Zhang, Li, Zhang, and Jin}]{zhang2024hirope}
Kechi Zhang, Ge~Li, Huangzhao Zhang, and Zhi Jin. 2024{\natexlab{a}}.
\newblock Hirope: Length extrapolation for code models using hierarchical position.
\newblock In \emph{Proceedings of the 62nd Annual Meeting of the Association for Computational Linguistics (Volume 1: Long Papers)}, pages 13615--13627.

\bibitem[{Zhang et~al.(2024{\natexlab{b}})Zhang, Li, Li, Shi, and Jin}]{zhang2024codeagent}
Kechi Zhang, Jia Li, Ge~Li, Xianjie Shi, and Zhi Jin. 2024{\natexlab{b}}.
\newblock Codeagent: Enhancing code generation with tool-integrated agent systems for real-world repo-level coding challenges.
\newblock \emph{arXiv preprint arXiv:2401.07339}.

\end{thebibliography}



\clearpage
\newpage
\appendix

\section{Rrecision-based results in RQ2}
\label{sec:appendix-rq2}

\begin{figure*}
\centering
\includegraphics[width=0.9\linewidth]{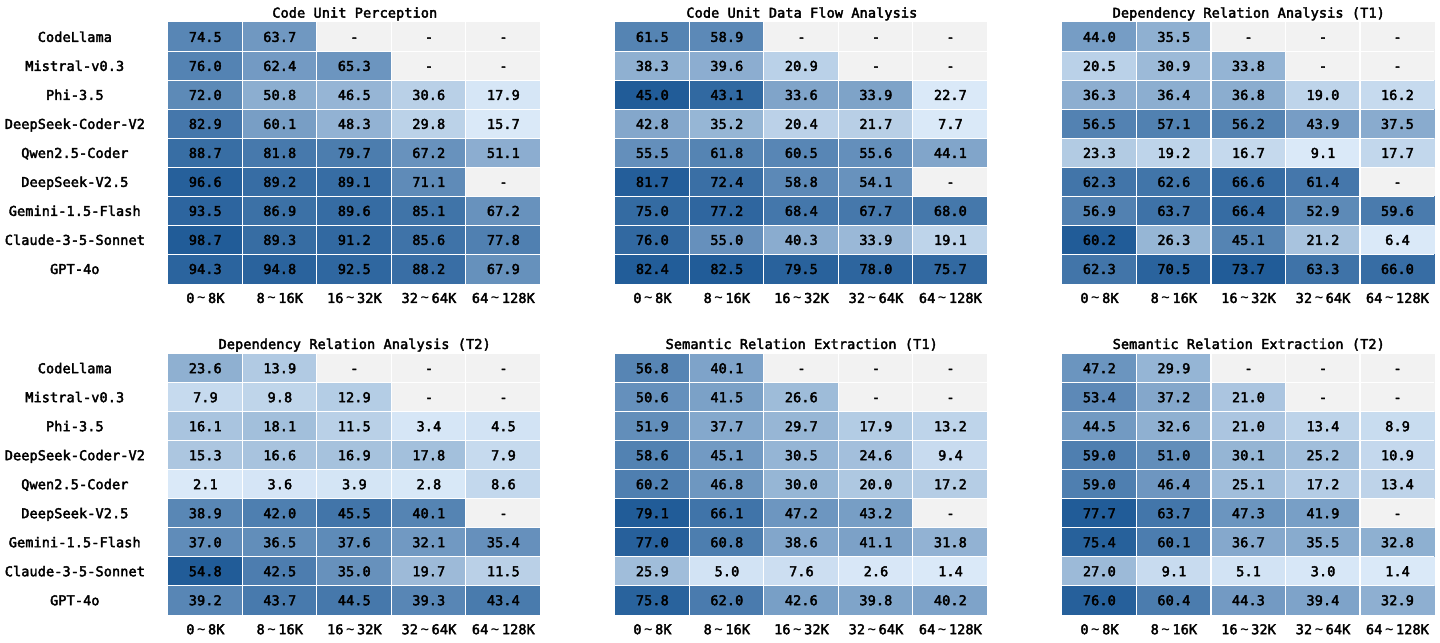}
\caption{Performance comparison across long code lengths on tasks which can be measured by precision-based metrics. (grey blocks indicate unavailable configurations). The rate of performance degradation exhibits task-specific and model-specific patterns. }
\label{Figure: results-precision}
\end{figure*}

We present the performance of LCLMs on several tasks which can be measured by precision-based metrics in Figure \ref{Figure: results-precision}.

\begin{figure}[t]
\centering
\includegraphics[width=0.8\columnwidth]{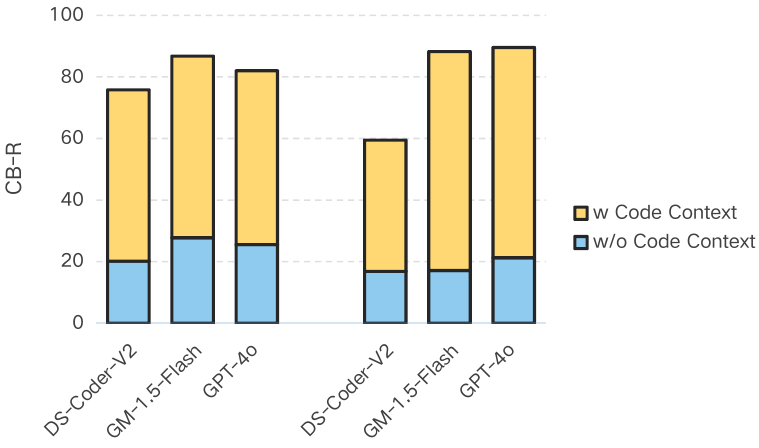}
\caption{Assessing long code understanding vs. memorization on CU\_SA (left) and DRA\_T2 (right) tasks.}
\label{figure: memory}
\end{figure}

\begin{figure}[t]
\centering
\includegraphics[width=0.8\columnwidth]{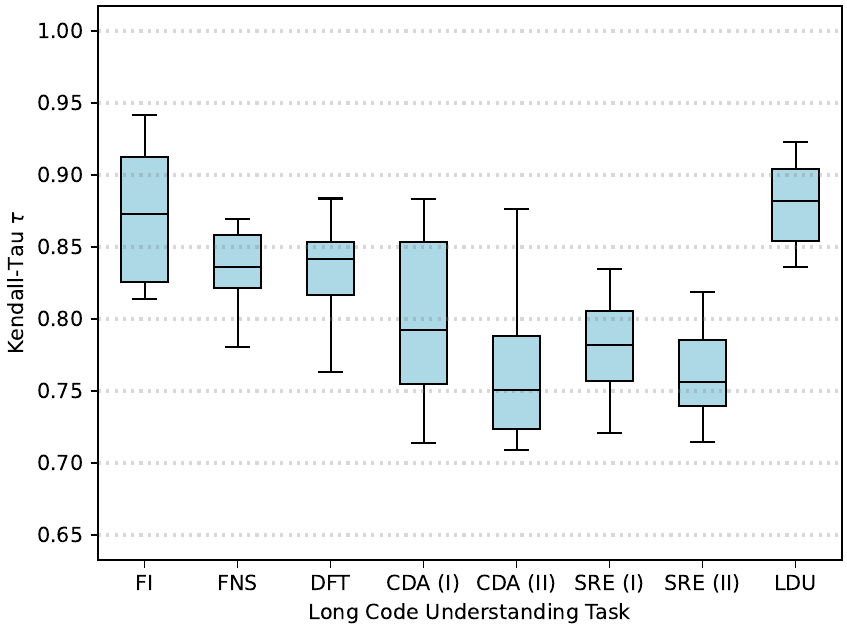}
\caption{The value of Kendall-Tau $\tau$ between our automatic metrics and human evaluation.}
\label{figure: Reliable Evaluation Metrics}
\end{figure}

\begin{figure*}[h]
\centering
\includegraphics[width=1.5\columnwidth]{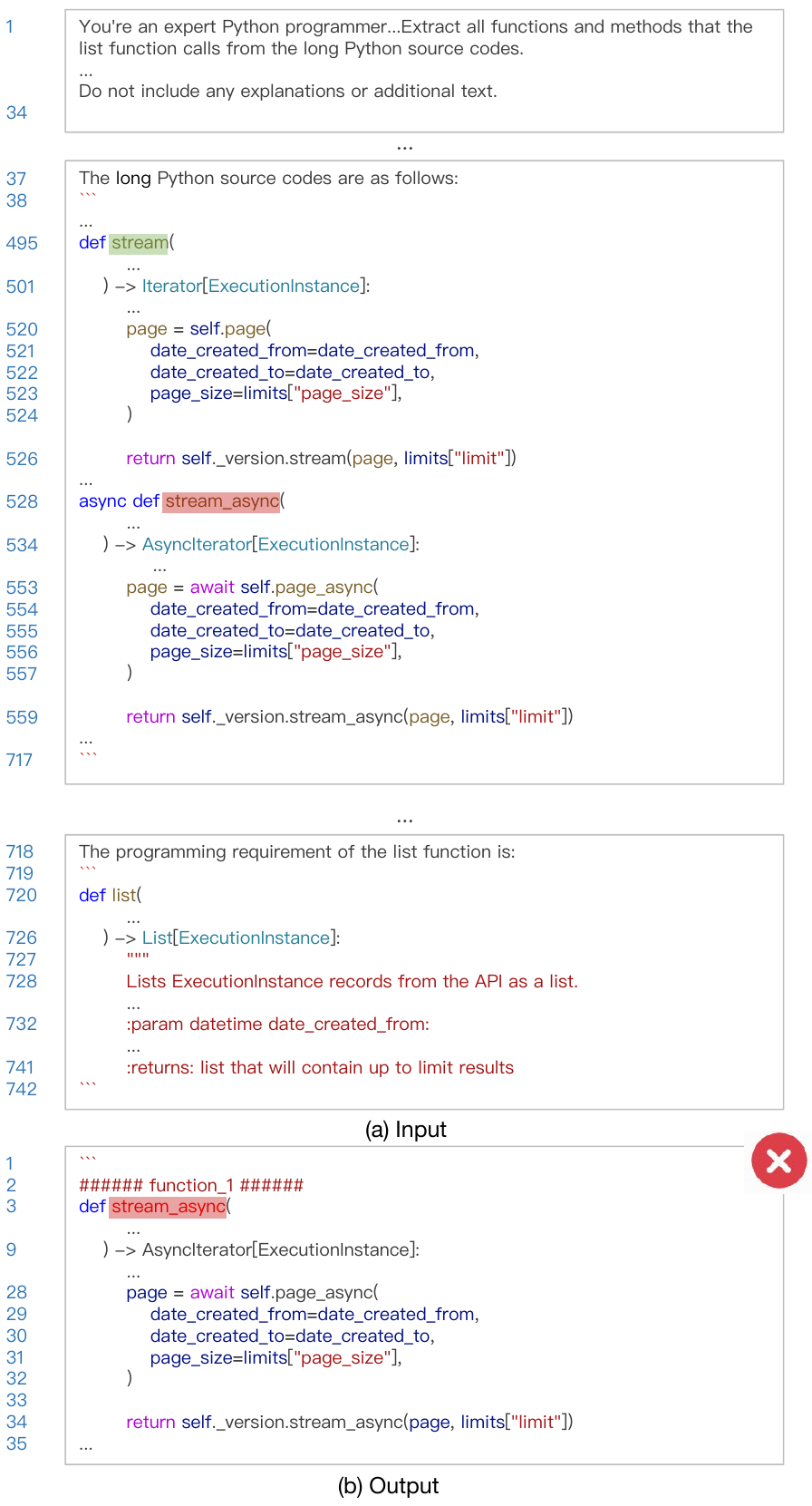}
\caption{For the dependency relation analysis task, the output of GPT-4o extracts a error code unit ``stream\_async" that is confusing to correct invoked function ``stream".}
\label{Figure: case1}
\end{figure*}

\begin{figure*}[h]
\centering
\includegraphics[width=1.5\columnwidth]{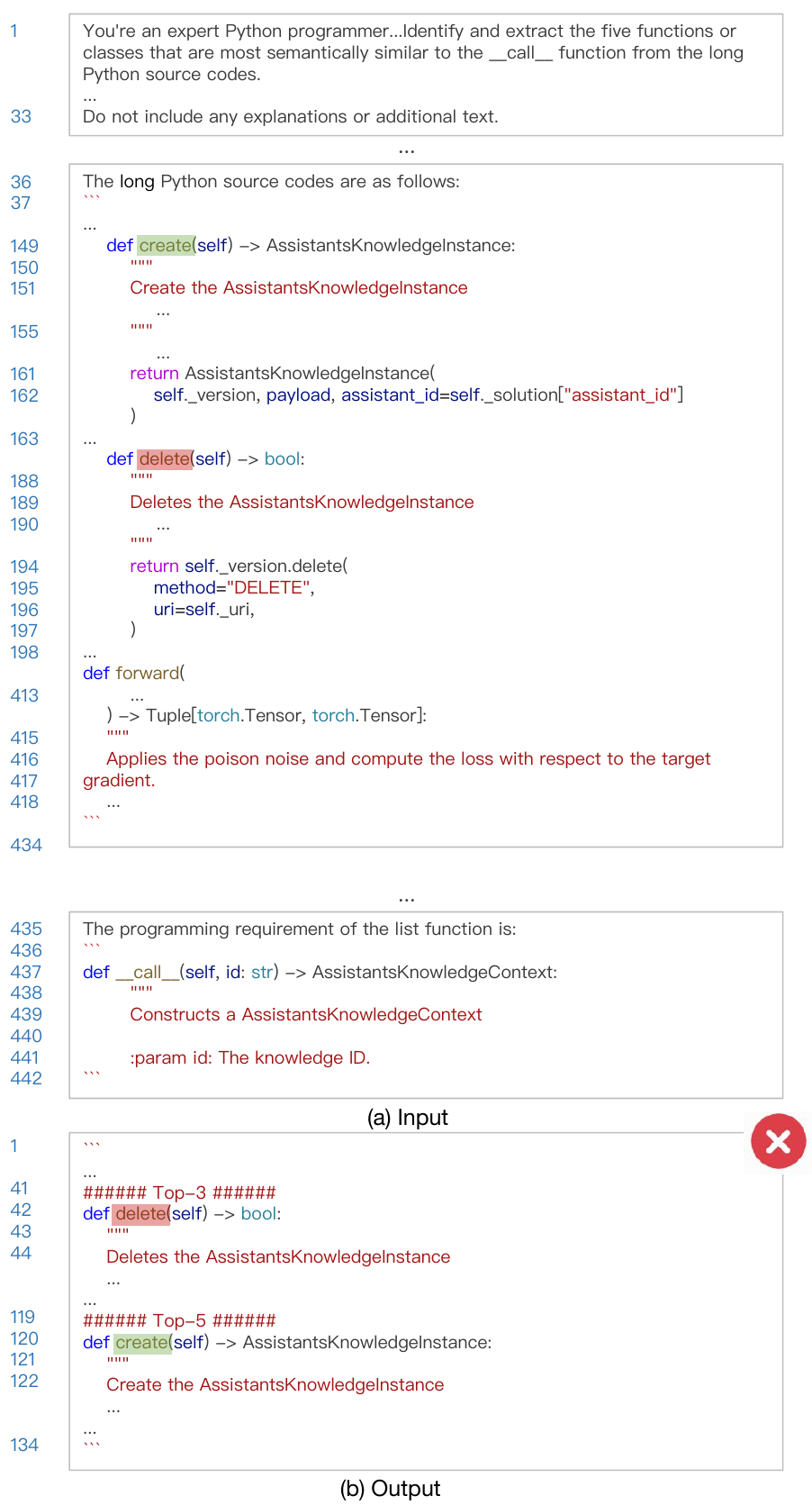}
\caption{For the semantic relation extraction task, the output contains an error ``delete" function which has opposite functionalities to the anchor input, \ie the given natural language description. }
\label{Figure: case2}
\end{figure*}

\section{Case Study}
\label{sec:appendix-case}

Figure \ref{Figure: case1} shows a representative error case in the dependency relation analysis task. We can find that the output of GPT-4o extracts an error code unit ``stream\_async" that is confusing to correct invoked function ``stream" since the two code snippets have similar structures.

Figure \ref{Figure: case2} demonstrates a generated result of GPT-4o in the semantic relation extraction task. We can observe that the output contains an error ``delete" function which has opposite functionalities to the anchor input, \ie the given natural language description.

\section{Instructions in Our Benchmark} 
\label{sec:appendix-instruction}

We present the instructions employed in diverse tasks. Each instruction has undergone iterative refinement to ensure that different models can not only achieve relatively high metrics but also produce outputs that appear satisfactory and are applicable to real-world development scenarios when using these instructions.
These instructions standardize the output format of the models (even though some models may not output strictly in accordance with these specifications), facilitating the parsing of the streaming output of the models using regular expressions during the evaluation process.

\begin{figure*}
\centering
\begin{AcademicBox} [(a) Code Unit Perception]
    \small
    \vspace{1mm}
    You're an expert Python programmer. Identify all defined Python functions and extract their function names from the long Python source codes.\\

    Only extract the Python function names that are defined in the long Python source codes.\\
    Do not extract the Python function names that are mentioned but not defined in the long Python source codes.\\
    Do not extract the Python function names that are defined in the classes of the long Python source codes.\\
    All extracted Python function names are stored in a list \*\*Function\_names\*\* according to the order these function names appear in the long Python source codes.\\

    Make sure the output is strictly in this format:\\
    ``\\
    Function\_names = [Function\_name\_1, Function\_name\_2, ..., Function\_name\_k]\\
   ''\\

    Do not include any explanations or additional text.\\
\end{AcademicBox}
\end{figure*}

\begin{figure*}
\centering
\begin{AcademicBox} [(b) Code Unit Data Flow Analysis]
    \small
    \vspace{1mm}
    You are an expert Python programmer. Identify the \{Target\_function\_name\} function from the long Python source code and extract the lines where the value of the \{Target\_variable\_name\} variable changes within the \{Target\_function\_name\} function.\\

    Only extract lines where the value of the \{Target\_variable\_name\} variable changes.\\
    Do not extract lines where the \{Target\_variable\_name\} variable is referenced but its value does not change.\\

    Return the extracted lines according to the order they appear in the \{Target\_function\_name\} function.\\

    Make sure the output is in the following format:\\

    ``\\
    \#\#\#\#\#\# The extracted code line \#\#\#\#\#\#\\
    \{extracted\_code\_line\}\\
    \#\#\#\#\#\# Line number \#\#\#\#\#\#\\
    \{line\_number\}\\

    \#\#\#\#\#\# The extracted code line \#\#\#\#\#\#\\
    \{extracted\_code\_line\}\\
    \#\#\#\#\#\# Line number \#\#\#\#\#\#\\
    \{line\_number\}\\
    ...\\
    \#\#\#\#\#\# The extracted code line \#\#\#\#\#\#\\
    \{extracted\_code\_line\}\\
    \#\#\#\#\#\# Line number \#\#\#\#\#\#\\
    \{line\_number\}\\

    ``\\

    Do not include any explanations or additional text in the output.\\
\end{AcademicBox}
\end{figure*}

\begin{figure*}
\centering
\begin{AcademicBox} [(c) Code Unit Semantic Analysis]
    \small
    \vspace{1mm}
    You're an expert Python programmer. Understand the long Python source codes. Identify and extract the function that satisfies the given programming requirement.\\

    Extract the entire source code of the function that satisfies the given programming requirement.\\
    Do not only extract the name of the function that satisfies the given programming requirement.\\

    Extracted the entire source code of the function:\\
    ``\\
    def function\_name(parameters):\\
        \# function body\\
    ''\\

    Do not include any explanations or additional text. \\
\end{AcademicBox}
\end{figure*}

\begin{figure*}
\centering
\begin{AcademicBox} [(d) Dependency Relation Analysis (T1)]
    \small
    \vspace{1mm}
    You're an expert Python programmer. Extract all functions and methods that the \{Target\_function\_name\} function calls from the long Python source codes.\\ 

    Only extract functions and methods from the given long Python context.\\
    Do not extract functions and methods from the Python standard library or third-party libraries.\\
    Only extract functions and methods that the \{Target\_function\_name\} function directly calls.\\
    Do not extract functions and methods that the \{Target\_function\_name\} function indirectly calls.\\
    Extract the entire source code of functions and methods that the \{Target\_function\_name\} calls.\\
    Do not only extract the name of functions and methods that the \{Target\_function\_name\} function calls.\\
    Ensure indentation is preserved.\\

    Please follow the format to complete the skeleton below:\\
    ``\\
    \#\#\#\#\#\# function\_1 \#\#\#\#\#\#\\
    def function\_name\_1(parameters):\\
        \# function body\\

    \#\#\#\#\#\# function\_2 \#\#\#\#\#\#\\
    def function\_name\_2(parameters):\\
        \# function body\\

    \#\#\#\#\#\# method\_1 \#\#\#\#\#\#\\
    def method\_name\_1(self, parameters):\\
        \# method body\\

    \#\#\#\#\#\# method\_2 \#\#\#\#\#\#\\
    def method\_name\_2(self, parameters):\\
        \# method body\\
    ''\\

    Do not include any explanations or additional text. \\
\end{AcademicBox}
\end{figure*}

\begin{figure*}
\centering
\begin{AcademicBox} [(e) Dependency Relation Analysis (T2)]
    \small
    You're an expert Python programmer. Understand the Python programming requirement of the \{Target\_function\_name\} function. Extract all functions and methods that the \{Target\_function\_name\} function calls from the long Python source codes.\\
    
    Only extract functions and methods from the given long Python context.\\
    Do not extract functions and methods from the Python standard library or third-party libraries.\\
    Only extract functions and methods that the \{Target\_function\_name\} function directly calls.\\
    Do not extract functions and methods that the \{Target\_function\_name\} function indirectly calls.\\
    Ensure indentation is preserved.\\
    
Do not include any explanations or additional text.\\ 
\end{AcademicBox}
\end{figure*}

\begin{figure*}
\centering
\begin{AcademicBox} [(f) Semantic Relation Extraction (T1)]
    \small
    \vspace{1mm}
    You're an expert Python programmer. Identify and extract the functions or classes that are most semantically similar to the \{Target\_function\_name\} function from the long Python source codes.\\

    Extract the entire source code of functions and classes that are the top-5 semantically similar to the \{Target\_function\_name\} function.\\
    Do not only extract the name of functions and classes that are the top-5 semantically similar to the \{Target\_function\_name\} function.\\
    The order of extracted five functions or classes is in order of decreasing similarity to the \{Target\_function\_name\} function.\\
    Ensure indentation is preserved.\\
    
    **Do not extract the target function \{Target\_function\_name\} itself.**\\
    
    Please follow the format to complete the skeleton below:\\
    ``\\
    \#\#\#\#\#\# Top-1 \#\#\#\#\#\#\\
    def name\_1(parameters):\\
        \# function body\\
    
    \#\#\#\#\#\# Top-2 \#\#\#\#\#\#\\
    def name\_2(parameters):\\
        \# body\\
    
    \#\#\#\#\#\# Top-3 \#\#\#\#\#\#\\
    def name\_3(parameters):\\
        \# body\\
    
    \#\#\#\#\#\# Top-4 \#\#\#\#\#\#\\
    def name\_4(parameters):\\
        \# body\\
    
    \#\#\#\#\#\# Top-5 \#\#\#\#\#\#\\
    def name\_5(parameters):\\
        \# body\\
    ''\\
    
    Do not include any explanations or additional text.\\
\end{AcademicBox}
\end{figure*}

\begin{figure*}
\centering
\begin{AcademicBox} [(g) Semantic Relation Extraction (T2)]
    \small
    \vspace{1mm}
    You're an expert Python programmer. Understand the Python programming requirement of the \{Target\_function\_name\} function. Identify and extract the functions or classes that are most semantically similar to the \{Target\_function\_name\} function from the long Python source codes. \\

    Extract the entire source code of functions and classes that are the top-5 semantically similar to the \{Target\_function\_name\} function.\\
    Do not only extract the name of functions and classes that are the top-5 semantically similar to the \{Target\_function\_name\} function.\\
    The order of extracted five functions or classes is in order of decreasing similarity to the \{Target\_function\_name\} function.\\
    Ensure indentation is preserved.\\
    
    **Do not extract the target function \{Target\_function\_name\} itself.**\\
    
    Please follow the format to complete the skeleton below:\\
    ``\\
    \#\#\#\#\#\# Top-1 \#\#\#\#\#\#\\
    def name\_1(parameters):\\
        \# function body\\
    
    \#\#\#\#\#\# Top-2 \#\#\#\#\#\#\\
    def name\_2(parameters):\\
        \# body\\
    
    \#\#\#\#\#\# Top-3 \#\#\#\#\#\#\\
    def name\_3(parameters):\\
        \# body\\
    
    \#\#\#\#\#\# Top-4 \#\#\#\#\#\#\\
    def name\_4(parameters):\\
        \# body\\
    
    \#\#\#\#\#\# Top-5 \#\#\#\#\#\#\\
    def name\_5(parameters):\\
        \# body\\
    ''\\
    
    Do not include any explanations or additional text.\\
\end{AcademicBox}
\end{figure*}

\begin{figure*}
\centering
\begin{AcademicBox} [(h) Long Documentation Understanding]
    \small
    \vspace{1mm}
    You are an expert Python programmer. Understand the multiple natural language documentations and identify the one that describes the \{Target\_API\_name\} API.\\

    Each documentation is labeled with a sequence number and enclosed between special markers: \textless BEGIN \textgreater indicates the start of a documentation, and \textless END \textgreater indicates its conclusion.\\
    
    Extract and return only the complete documentation corresponding to the \{Target\_API\_name\} API, exactly as it appears. \\
    
    Please follow the format to complete the skeleton below:\\
    
    ``\\
    \# The sequence number\\
    \{sequence number\}\\
    
    \# The complete documentation\\
    \{complete documentation\}\\
    ''\\
    
    Do not include any explanations or additional text.\\
\end{AcademicBox}
\end{figure*}

\end{document}